\documentclass[water,article,accept,oneauthors,10pt,a4paper]{mdpi} 

\usepackage{booktabs}
\usepackage{multirow}
\usepackage{soul}
\usepackage{microtype}

\newcommand \be{\begin{eqnarray}}
\newcommand \ee{\end{eqnarray}}
\newcommand \V{\vec}
\newcommand \ba{\begin{align}}
\newcommand \eea{\end{align}}

\firstpage{1} 
\articlenumber{x}
\doinum{10.3390/------}
\pubvolume{xx}
\pubyear{2017}
\copyrightyear{2017}
\externaleditor{Academic Editors: Elmar Christof Fuchs,  Jakob Woisetschläger, Adam D. Wexler and Astrid H. Paulitsch-Fuchs}
\history{Received: 24 March 2017; Accepted: 15 May 2017; Published: date}

\pdfoutput=1

\Title{Reversed Currents in Charged Liquid Bridges}

\Author{Klaus Morawetz $^{1,2,3}$}

\address{%
$^{1}$ \quad M\"unster University of Applied Sciences,
Stegerwaldstrasse 39, 48565 Steinfurt, Germany; morawetz@fh-muenster.de; Tel.:+49 2551 962 411\\
$^{2}$ \quad International Institute of Physics (IIP),
Av. Odilon Gomes de Lima 1722, 59078-400 Natal, Brazil\\
$^{3}$ \quad Max-Planck-Institute for the Physics of Complex Systems, 01187 Dresden, Germany}
\corres{Correspondence: morawetz@fh-muenster.de}

\abstract{The velocity profile in a water bridge is reanalyzed. Assuming
  hypothetically that the bulk charge has a radial distribution, a surface potential is formed that is analogous to the Zeta
  potential. The~Navier--Stokes equation is solved, neglecting the convective term;
then, analytically and for special field and potential ranges, a sign change of
the total mass flow is reported caused by the radial charge~distribution.
}

\keyword{classical transport; electrokinetic effects;
electrohydrodynamics; electrorheological fluids; Navier--Stokes equations
}

\PACS{05.60.Cd; 
47.57.jd;
47.65.-d; 
83.80.Gv
}

\begin{document}



\section{Introduction}

Although it has been known for over 100 years \cite{arm1893}, the formation of a water bridge between two beakers under high voltage has not lost its fascination.  It is interesting to understand the underlying electrohydrodynamics of such water bridges at nanoscale since liquid bridging is not restricted to water but can also be observed
in other liquids \cite{Sa10} and is characteristic of polar dielectric liquids \cite{WLSWF14}. Therefore, its cause might be in electrohydrodynamics \cite{Me69} rather than in molecular-specific structures. 

Applications can be found in atomic force microscopy \cite{Sa06};
phenomena in micro-fluidics \cite{Sq05}; up to electrowetting problems \cite{Ju10} to confine chemical reactions \cite{GM06}. 
Molecular dynamical simulations \cite{CHVWS10} help to understand the underlying mechanism, especially what leads a water filament to overcome the surface tension \cite{CZG08}. 
The formation of a water bridge across the lipid bilayer is considered as the
first stage of electroporation for the intrusion of individual water molecules
into the membrane interior~\cite{HLV13}. The dynamics of charged liquids is also used for capillary jets \cite{Ga97}. Some current applications include ink printers and electrosprays \cite{Ga10,Hi10} and the breakup dynamics of free surfaces and flows~\cite{E93,Egg97}.

Such water bridges can even be stretched and used for flexible nanofluids \cite{CWWWG16}. Water bridges exhibit viscoelastic behavior with the possibility to measure its Young modulus \cite{TSV13}. There is an interplay between field-induced polarization, surface tension, and
condensation \cite{GS03,CZG08} such that the forces of dielectric and surface tensions simultaneously hold the bridge against gravity \cite{NLAJI13}.
For an overview of the different forces
occurring in microelectrode structures, see \cite{Ra98}.

In the absence of bulk charges, the forces on the water stream are caused by the
pressure of the polarizability of water due to the high dielectric susceptibility
$\epsilon$
. This pressure leads to the catenary form of the water bridge---like a hanging chain \cite{WSSSS09}. While the simplified model of \cite{Sa10}, employing a capacitor picture, already leads to a critical field strength for the formation of
the water bridge, the~catenary model \cite{WSSSS09} or a charged catenary \cite{M12} allows the determination of the static and dynamical stability conditions where charged
liquid bridges are possible. The creeping height, the~bridge radius and length
as well as the shape of the bridge were calculated, showing an asymmetric
profile in agreement with observations \cite{Wo10,ML10}. The flow profile was
obtained from the Navier--Stokes equation leading to a mean velocity which
combines charge transport with neutral mass flow \cite{M12a}. 
It shows that even the uncharged catenary
provides a minimal critical field strength for the water bridge
formation in dependence on the length of the bridge. This critical field
strength is modified if charges are present in the bridge \cite{M12}. The
occurrence of charges due to dissociation has also been used to explain the
I--V hysteresis reported in \cite{OFRF14}.

In this paper, it is assumed that the charge carriers play an important
role, such that surface charges might be formed. This hypothetical assumption
is analyzed and leads to a surface potential which essentially influences the charge
transport. Since both mass and charge transport are coupled, here we
find the possibility that the mass transport might show a reverse
behavior dependent on the parameters. This study is motivated by the 
bidirectional flow visualized recently in \cite{WDFWRFR17} and measured with mass and charge transfer in \cite{FAEW10}.

\section{Velocity Profile with Bulk Charges}

The link between the Ohm picture of 
a constant current density of charged particles and the dependence of the mass
flow on the cross-area  for incompressible fluids has been found in a combined
flow expression \cite{M12a}. Possible charges with density $\rho_c$ in water
move according to the applied electric field $E$ and will create a mean mass motion of density $\rho$. Such access charge has been found in \cite{TSGF16,FSWKW16}.
Let us first recall the main steps for such a transport picture and then investigate how a radial charge distribution changes the results.

In the stationary case and neglecting the nonlinear convection, the Navier--Stokes equation~\cite{Ch94}~reads
\be
\eta \nabla^2{\V v}-\nabla P+\rho_c {\V E}=0
\label{Navier}
\ee
with the gradient of the electric pressure given in the direction of the bridge of length $L$ by
\be
-\nabla P={\epsilon_0 (\epsilon-1) E^2\over 2 L}={b(E)\over 2 L}\rho g
\label{gradP}
\ee
with the dielectric constant $\epsilon$ and introducing the creeping height
\be
b(E)={\epsilon_0 (\epsilon-1) E^2\over \rho g}.
\ee

For details about the gradient of the electric pressure, see Equations (8) and (22) of \cite{M12a}. 
This leads from (\ref{Navier}) to the equation for the radial-dependent velocity
\be
{\eta\over \rho g} {d\over d r} \left (r {d v \over dr}\right )+ r \left
  ({b\over 2 L}+c\right )=0
\label{diffg}
\ee
where we have introduced the dimensionless ratio $c$ of the force density on the
charges by the field, $\rho_cE$, to the gravitational force density, $\rho g$, in the form 
\be
c(\rho_c,E)={\rho_c E\over \rho g}.
\ee 

The solution of (\ref{diffg}) provides the velocity profile in the direction of the bridge as
\be
v(r)-v(R)=v_0 \left ({b\over 2 L}+c\right ) \left (1-{r^2\over R^2}\right ) 
\label{ur}
\ee
where $R$ is the radius of the bridge and the
characteristic velocity reads
\be
v_0={\rho g R^2 \over 4 \eta}.
\label{u0}
\ee

In \cite{M12a}, it was assumed that the undetermined
velocity at the surface of the bridge $v(R)$ vanishes. 

\section{Surface Potential}

The direction of the flux not only depends on the applied electrostatic
field but also on the relative electronegativity of the electrodes
\cite{GELGBG15}. A complex bi-directional mass transfer pattern has been found
in such bridges \cite{WLSWF14,WDFWRFR17}. Especially in the latter paper, the
bidirectional flow was observed by charged isotopes. The light water flows
above the heavier water. Thus, it is justified to assume bulk charges.
Already in \cite{Me69}, it was considered that the bulk charges might be realized in a surface sheet. Ionized charges at the anode migrate to the outer surface of the bridge \cite{TSGF16}. Therefore, the migration of
charges to the surface should be considered to be 
forming a charged surface sheet which will be an extension of \cite{M12a}. Now, let~us develop step by step the idea of a surface charge and the corresponding potential.

The force on an ionic charge $e_i$ in an electric field with small velocities $v_i$ can be well approximated by the Stokes force
\be
\V F=e_i \V E=6 \pi \eta r_i \V v_i
\ee
where $r_i$ is the radius and $\eta$ the viscosity. With the conductivity $\sigma$, the $\zeta$ potential can be
defined from the velocity 
\be
\V v_i={e_i\over 6 \pi \eta r_i} \V E={\sigma\over \rho_i} \V E=-\epsilon \epsilon_0 {\zeta\over \eta} \V E
\ee
and can be written in two forms
\be
\zeta=-{e_i\over 6 \pi \epsilon \epsilon_0 r_i}=-{\sigma \eta\over \epsilon \epsilon_0\rho_i}.
\label{zeta}
\ee

This $\zeta$ potential actually describes the electric potential at the
surface of the bridge which we will now see from the analogous idea as one
describes electro-osmose. We assume that the charge density in the bridge
consists of a homogeneous bulk charge $\rho_b$,
\be 
\rho_c=\rho_b+\rho_r(r),
\ee
and a radial-dependent modulation of the charge density
$\rho_r(r)$ according to a distribution that is analogous to the screening cloud. This is a purely hypothetical assumption which we will justify by
the observation that it leads exactly to the finite potential $\zeta$ at the surface. In other
words, in the case that the surface potential of the bridge is at the boundary
between liquid and air, the following considerations are 
the result. 
The Poisson equation for the electrostatic potential $\Psi$ in that case reads
\be
\nabla^2\Psi =-{\rho_r\over \epsilon\epsilon_0}=-{1\over
  \epsilon\epsilon_0}\sum\limits_i n_i e_i \left ({\rm e}^{-e_i
    \Psi/T}-1\right )\approx \kappa^2 \Psi
\label{Pois}
\ee
with the squared inverse screening length
$
\kappa^2=\sum\limits_i {n_i e_i^2 \over \epsilon\epsilon_0 T}$. Here, we have
assumed that the integrated spatial inhomogeneous charge density modulation 
is zero such that besides the homogeneous $\rho_b$ no total access charge 
remains.
The Poisson Equation (\ref{Pois}) is readily solved 
for a radial dependence
\be
\Psi(r)=\zeta {I_0(\kappa r)\over I_0(\kappa R)}
\ee
with the potential at the surface $\zeta=\Psi(R)$ and the Bessel function $I_0$.
From (\ref{Pois}), we obtain the radial dependence of the spatial modulation of
the charge density
\be
\rho_r(r)=-\epsilon\epsilon_0 \nabla^2\Psi=-\epsilon\epsilon_0\kappa^2\Psi(r)
\ee
from which the additional body force follows
\be
\V f(r)=\rho_r(r) \V E=-\epsilon\epsilon_0\kappa^2\zeta {I_0(\kappa r)\over I_0(\kappa R)}\V E.
\label{fden}
\ee

\section{Velocity Profile with Bulk and Surface Charges}

Now we are ready to solve the Navier--Stokes Equation (\ref{Navier}) including 
this additional force density~(\ref{fden})
\be
\eta \Delta \V v- \V \nabla P+\V f+\rho_b \V E=0.
\label{NS}
\ee 

The bridge extends in the $x$-direction with the following geometry
\be
\V E=(E,0,0): \quad \V v=(v(r),0,0)
\ee
as well as a constant pressure gradient $\nabla P=\partial_x P$ of (\ref{gradP})
This leads to a simple equation for the radial dependence that is analogous to (\ref{diffg}) 
\be
{\eta\over r} {d\over d r} \left (r {d v \over dr}\right )+ \rho g \left ({b\over 2 L}+c\right )=\epsilon\epsilon_0 \kappa^2\zeta E {I_0(\kappa r)\over I_0(\kappa R)}.
\label{diffgc} 
\ee
Demanding that the solution is finite at $r=0$, one can integrate (\ref{diffgc}) and instead of~(\ref{ur}), we obtain
\ba
{v(r)\!-\!v(R)\!\over v_0}=
\left (\!{b\over 2 L}\!+\!c\right ) \left (\!1\!-\!{r^2\over R^2}\!\right )
\!+\!{4 \epsilon\epsilon_0\zeta E\over \rho g R^2}\left [\!{I_0(\kappa r)\over I_0(\kappa R)}\!-\!1\!\right ].
\label{sol}
\end{align}
This gives the velocity profile in dependence on the radius if a surface potential $\zeta$ is present. The~first part without such potential, (\ref{ur}), 
has been obtained in \cite{M12a} considering the bulk charges and the dielectric pressure due to the electric field. The second term is the new contribution arising from the surface charge.


The total volume flow relative to the surface flow 
can be calculated from (\ref{sol}) as
\be
{J(R)}&=&{2 \pi} \int\limits_0^R \left [v(r)-v(R)\right ] r dr= J_0\left [(\kappa R)^2-8 {\zeta\over \zeta_0}{I_2(\kappa R)\over I_0(\kappa R)}\right ]
\label{curr}
\ee
with 
\be
J_0&=&{\pi v_0\over 2 \kappa^2}\left ( {b\over 2 L}+c\right )\nonumber\\
\zeta_0&=&\left ({b\over 2 L}+c\right ){\rho g\over \epsilon\epsilon_0 E\kappa^2}
={(\epsilon-1) E\over 2 \epsilon \kappa^2 L}+{\rho_b\over \epsilon \epsilon_0 \kappa^2}.
\label{I0}
\ee

\begin{figure}[H]
\centerline{\includegraphics[width=8cm]{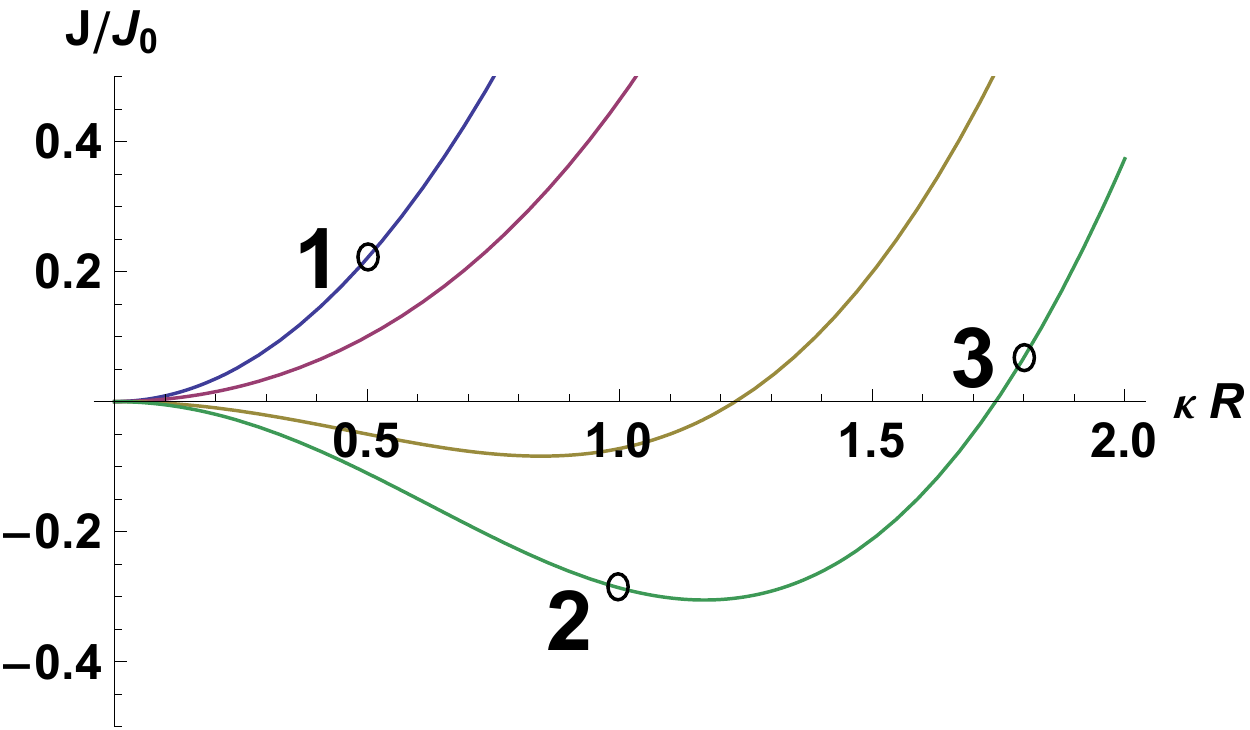}}
\caption{The total mass flow (\ref{curr}) vs. the dimensionless radius of the bridge. The parameter $8 \zeta/ \zeta_0$~=~$1,5,10,12$ from left to right. The numbered points represent the regions for which the velocity profile is given in Figure \protect\ref{zeta3}. \label{zeta1}}  
\end{figure}

\begin{figure}[H]
\centerline{\includegraphics[width=8cm]{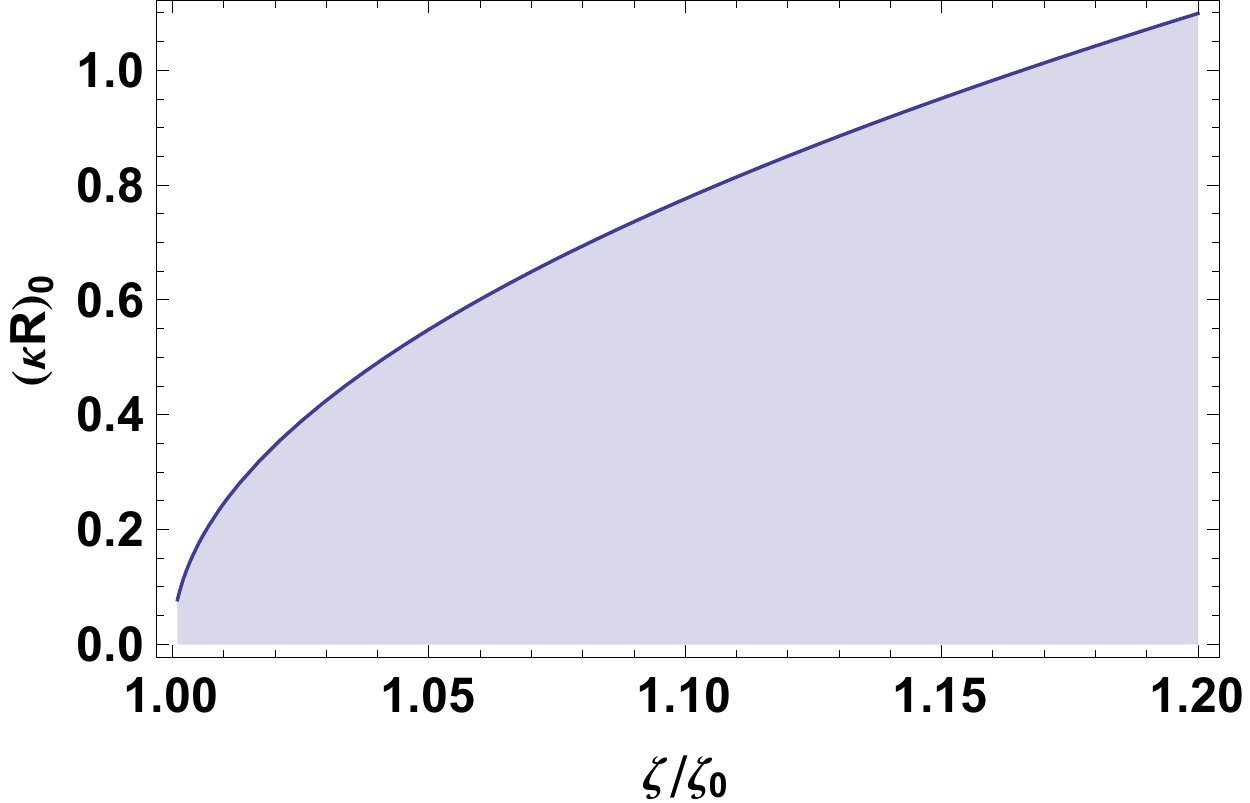}}
\caption{The critical parameter $(\kappa R)_0$ where the mass flow changes the sign vs. the $\zeta$ potential. The~shaded area is the region where the flow reverses the sign. \label{zeta2}} 
\end{figure}

The result is presented in Figure~\ref{zeta1} and one can see that the flow can change the direction if the $\zeta$ potential exceeds a critical value. This value can be 
found by observing in Figure~\ref{zeta1} that the sign change appears for small
$\kappa R$ leading to a zero in $J(R)$ at $(\kappa R)_0$. Therefore, from an expansion of (\ref{curr}) for $\kappa R$, the sign change can appear for
\be
{\zeta}>\zeta_0
\ee
with (\ref{zeta}) and (\ref{I0}).
The zero nonlinear $(\kappa R)_0$ in dependence on the $\zeta$ potential is plotted in
Figure \ref{zeta2}. The region of inverse flow is given for values of $\kappa R$ which are smaller than $(\kappa R)_0$ indicated by the shaded area.

\vspace{-6pt}

\begin{figure}[H]
\centerline{\includegraphics[width=8cm]{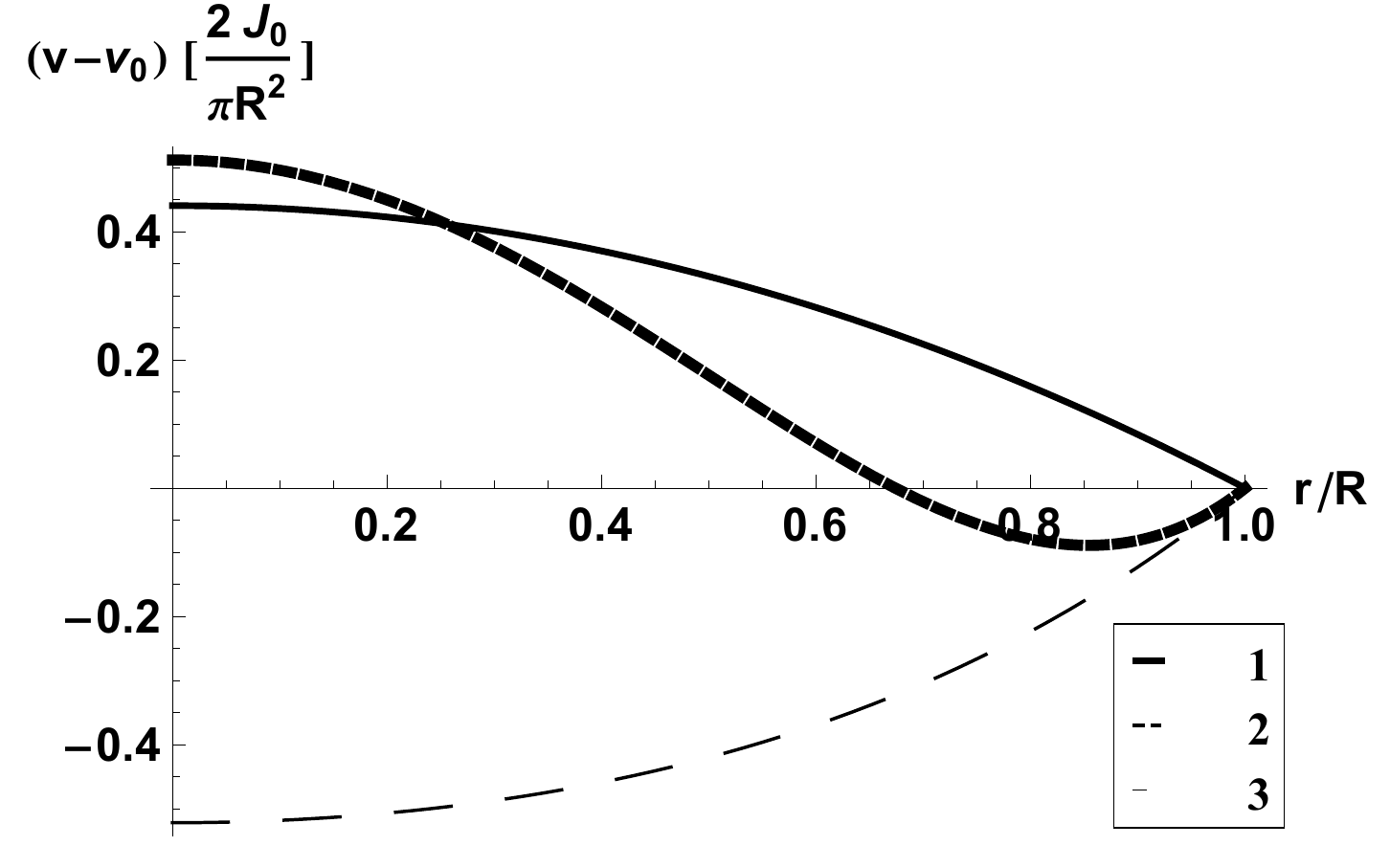}}
\caption{The radial velocity profile for the three numbered points of Figure \protect\ref{zeta1}. }
\label{zeta3}
\end{figure}

With the help of (\ref{I0}), the radial velocity profile (\ref{sol}) can be recast into the form
\ba
v(r)\!-\!v(R)\!=\!{2 J_0\over \pi R^2} \!\left \{ 
\!(\kappa R)^2 \left (\!1\!-\!{r^2\over R^2}\!\right )
\!+\!4 {\zeta\over \zeta_0} 
\left [
{I_0(\kappa r)\over I_0(\kappa R)}\!-\!1
\right ]
\!\right\}
\end{align}
and is presented in Figure \ref{zeta3}. We plot the radial velocity profile of
the three numbered points in Figure~\ref{zeta1}. One sees that point 3 of
Figure \ref{zeta1} leads to a sign change in the profile such that the total
mass flow has an opposite sign. Of course, this is only the ideal case when the
velocity at the surface vanishes $v(R)=0$. Otherwise, we have to add the value $\pi R^2 v(R)$ to the
mass flow (\ref{curr}), which we assume to be~small.

\section{Summary}
As motivated by the recent visualizations of bidirectional flow
\cite{WDFWRFR17}, we have considered, in addition to bulk charges, a spatial modulation of
the radial charge distribution such that a surface potential
occurs. Solving the Navier--Stokes equation leads to
a modified mass flow through the bridge. Dependent on the parameters, one can
find a range where the flow is changing its direction. The observation here is
that a surface potential can cause a line of zero velocity, i.e., a stagnation
line of mass flow, and can lead to a sign reversal of the total mass current. One will most probably observe bi-directional mass transfer patterns. The surface potential
can be thought of as being created by polar water molecules and charges separated
from neutral air. In order to keep the total charge density constant, such a
surface potential could balance the access charges. This is so far a purely hypothetical scenario.

\vspace{12pt}
\conflictsofinterest{The authors declare no conflict of interest.}

\bibliographystyle{mdpi}

\renewcommand\bibname{References}



\end{document}